\documentclass{emulateapj}
\usepackage{apjfonts}
\usepackage{natbib}
\usepackage{graphicx}
\usepackage{amssymb}

\def\msun{{~M}_{\odot}}

\def\be{\begin{equation}}
\def\ee{\end{equation}}

\begin{document}
\shorttitle{Episodic Jets as the Central Engine of
Gamma-Ray Bursts} \shortauthors{F. Yuan \& B. Zhang}

\title{Episodic Jets as the Central Engine of Gamma-Ray Bursts}

\author{Feng Yuan\altaffilmark{1} and
Bing Zhang\altaffilmark{2}}

\altaffiltext{1}{Shanghai Astronomical Observatory, Chinese Academy
of Sciences, 80 Nandan Road, Shanghai 200030, China;
fyuan@shao.ac.cn} \altaffiltext{2}{Department of Physics and
Astronomy, University of Nevada Las Vegas, Las Vegas, NV 89154, USA;
zhang@physics.unlv.edu}

\begin{abstract}
Most Gamma-ray bursts (GRBs) have erratic light curves, which demand
that the GRB central engine launches an episodic outflow. Recent
Fermi observations of some GRBs indicate a lack of the thermal
photosphere component as predicted by the baryonic fireball model,
which suggests a magnetic origin of GRBs. In view that powerful
episodic jets have been observed along with continuous jets in other
astrophysical black hole systems, here we propose an intrinsically
episodic, magnetically-dominated jet model for GRB central engine.
Accumulation and eruption of free magnetic energy in the corona of a
differentially-rotating, turbulent accretion flow around a
hyperaccreting black hole lead to ejections of episodic,
magnetically dominated plasma blobs. These blobs are accelerated
magnetically, collide with each other at large radii, trigger rapid
magnetic reconnection and turbulence, efficient particle
acceleration and radiation, and power the observed episodic prompt
gamma-ray emission from GRBs.

\end{abstract}

\section{Introduction}

Observations of gamma-ray bursts (GRBs) suggest that the GRB central engine
is able to launch an ultra-luminous, highly relativistic jet.
Most GRBs have erratic, rapidly varying
lightcurves \citep{fishman95}, typically lasting 10s to 100s of
seconds for long-duration GRBs, and less than 2 seconds for
short-duration GRBs. Recent observations of GRBs pose some
important constraints on the models of GRB central engine. First,
recent Fermi Large Area Telescope (LAT) observations suggest that most
GRBs have featureless smoothly joint broken power-law spectra (i.e.
Band-function spectra \citep{band93}) in a wide energy band
\citep{abdo09,zhang11}. The standard fireball model predicts
a strong thermal emission component from the fireball photosphere
\citep{pac86,meszaros00,peer06}. The non-detection likely suggests that the GRB outflow
is magnetically dominated \citep{zhang09,fan10}. Second, data analysis suggests
that the GRB lightcurves not only can be decomposed into many pulses
\citep{norris96,hakkila03}, but most of them can be also decomposed into the
superposition of a fast, spiky component
and a slow, smooth component \citep{gao12,vetere06}. A radiation model that
invokes magnetic turbulent reconnection triggered by collisions of
magnetically-dominated blobs has been proposed \citep{zhang10},
which can interpret the new observational data. This radiation model
requires a central engine that can eject an episodic, magnetically
dominated jet.

The leading model of GRB central engine invokes a
hyper-accreting black hole \citep{narayan92,meszaros99,narayan01}. In most previous
GRB central engine models, the rapid variability in the erratic light
curves is attributed either to the intermittency of the accretion
flow, i.e. a time-dependent accretion rate $\dot M$
\citep{macfadyen99}, or to the instability during the
propagation of the jet in the stellar envelope
\citep{zhang03,morsony10}.
For a magnetically dominated central engine, the leading model is the
Blandford-Znajek (BZ) mechanism \citep{blandford77}, which
requires a large-scale open magnetic field connecting the black hole
and an external astrophysical load, whose origin
is an open question. This model also
tends to generate a continuous jet, unless the accretion rate
is highly variable. Screw or kink instabilities
\citep{li00,mizuno09} are invoked to disrupt a continuous jet and produce discrete blobs.
Since the BZ mechanism is powered by accretion, the immediate
advantage of involving BH spin rather than accretion disk as the source
of jet power is not evident.

On the other hand, in addition to the continuous jets, episodic jets, which is intermittent and in the form of discrete moving
blobs, have been observed in other black hole systems. A magnetohydrodynamical model has
been proposed to explain the formation of these episodic jets
\citep{yuan09}. In this paper, we propose a central engine model for
GRBs, based on this idea. A review on observations and the model of
episodic jets are presented in Section 2. Our model is delineated in
detail in Section 3. The salient features of the model are
summarized in Section 4.

\section{Episodic Jets in Black Hole Systems and Their Formation Mechanism}

Episodic jets are most evidently observed in black hole X-ray binaries, e.g., in GRS 1915+105 and GRO J1655-40 \citep{mirabel94,hjellming95,fender2004},
and also in AGNs, as manifested as knots or blobs,
e.g., in 3C 120 \citep{marscher02,chat09} and NGC~4258 \citep{doi2011}.
In the case of 3C~120, knots have been directly resolved as due to episodic ejections and their ejection times have been determined by extensive radio monitoring observations. In the case of the supermassive black hole in our Galactic center, we have strong
evidence for episodic jets, but no continuous jets have ever been
detected \citep{yuan11}. The differences of the properties of the two types
of jets are summarized in \citet{fender2004}. Among other things, episodic
jets are much faster and more powerful. The collision between blobs is often
invoked to explain the flares detected in AGN jets.

\citet{yuan09} proposed a magnetohydrodynamical model for the
formation of episodic jets, by analogy of coronal
mass ejection (CME) in the Sun. The basic scenario is the following.
Closed magnetic field lines continuously emerge out of the accretion
flow into the corona. Because of shear and turbulent motion of the
accretion flow, the field line is twisted and deformed, resulting in
formation of a flux rope in the corona. The flux rope is
initially in force balance between magnetic tension and magnetic
compression forces. Energy and helicity are accumulated and stored
until a threshold is reached. The system then loses its equilibrium
and the flux rope is thrust outward by the magnetic compression
force in a catastrophic way, which causes an episodic jet.
After a magnetic blob is ejected, the magnetic tension is temporarily
relaxed. Later magnetic field emergence and distortion restart, and
the above delineated process repeats. Within this scenario,
no large-scale open magnetic field lines are needed. The basic picture
has gained supports from some numerical simulations \citep*[e.g.,][]{romanova98,kudoh02,machida04}.

Within the GRB context, several magnetic central engine models have
been proposed along the similar line. In the context of their neutron
star central engine model, \citet{blackman96} pointed out that magnetically dominated blobs rather than a continuous jet have the advantage in reproducing the time structure of GRBs. \citet{kluzniak98} and \citet{dai98} also emphasized the importance of accumulation of magnetic fields and subsequent emergence to account for an episodic jet. A similar idea was proposed to interpret X-ray flares \citep{dai06}
in some short GRBs. \citet{uzdensky2006,uzdensky2007}
proposed a magnetic tower model for GRBs, which invokes extraction
of the rotational energy of an accretion disk by amplification of
the toroidal magnetic field \citep{Lyndenbell1996}. The model is
essentially time-independent, different from the model discussed
in \citet{yuan09} and below.

\section{Episodic Magnetic Jets in GRBs}

\begin{figure}
\epsscale{0.8}\plotone{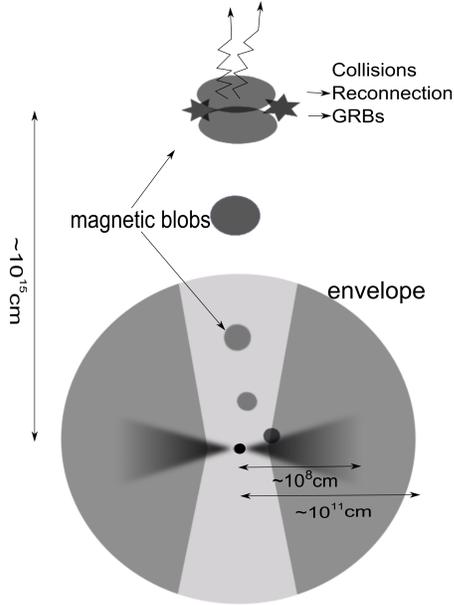} \epsscale{0.2} \caption{A cartoon
picture of the episodic jet model. The typical distances are marked.
} \label{fig:cartoon}
\end{figure}

We now extend the \citet{yuan09} scenario to GRBs. Our model can be sketched by Figure \ref{fig:cartoon}.
We consider a stellar-size black hole surrounded by a hyperaccretion accretion flow with a typical accretion rate $\dot{M}\sim 0.1 \msun {\rm s}^{-1}$. Neutrino cooling is important at $\lesssim 200
GM/c^2$ (neutrino-dominated accretion flow: NDAF); outside of this
radius the accretion flow is advection dominated \cite{chen07}
(ADAF). The solutions for NDAF and ADAF can be described as
\citep{narayan01,belo03}: \be \rho=6\times
10^{13}\dot{M}_{32}\alpha_{-2}^{-1.3}M_3^{-1.7}r^{-2.55}{\rm g~
cm}^{-3},\ee \be T_c=3\times
10^{10}\alpha_{-2}^{0.2}M_3^{-0.2}r^{-0.3}{\rm K},\ee
\be v_k=2\times 10^{10}r^{-0.5}{\rm cm\,s}^{-1},\ee \be v_r=2\times
10^{6}\alpha_{-2}^{1.2}M_3^{-0.2} r^{0.2}{\rm cm~s^{-1}},\ee and
\be \rho=3\times
10^{11}\alpha_{-2}^{-1}\dot{M}_{32}M_3^{-2}r^{-1.5}{\rm g\, cm}^{-3},\ee
\be T_c=3\times
10^{11}\alpha_{-2}^{-1/4}\dot{M}_{32}^{1/4}M_3^{-0.5}r^{-5/8}{\rm
K},\ee \be v_k=2\times 10^{10}r^{-0.5}{\rm cm\,s}^{-1},\ee
\be v_r=\frac{\alpha}{\sqrt{5}}R\Omega_k=10^{8}\alpha_{-2} r^{-0.5}{\rm
cm~s^{-1}},\ee respectively. Here $\rho$ is the density, $T_c$ is the
temperature at the equatorial plane, $v_k=\sqrt{GM/R}$
and $v_r$ are the Keplerian
and radial velocities, $\alpha=0.01\alpha_{-2}$ is the viscous
parameter, $r$ is radius $R$ in unit of $2GM/c^2$,
$\dot{M}_{32}$ is the mass accretion rate in
unit of $10^{32} {\rm g\,s}^{-1}$, and $M_3$ is the black hole mass
in unit of $3\msun$. The viscous timescales of the inner NDAF and
outer ADAF are then \be t_{\rm vis,NDAF}=R/v_r=17
\alpha_{-2}^{-1.2}M_3^{1.2}(r/100)^{0.8}~{\rm s},\ee and \be t_{\rm
vis,ADAF}=R/v_r= 30 \alpha_{-2}^{-1}M_3(r/200)^{1.5}~{\rm s},
\label{tvis} \ee respectively. Hereafter we normalize the NDAF
solutions at $r = 100$ where the disk solution transits from NDAF to
ADAF. All our ADAF solutions are normalized to $r=200$, which
corresponds to a viscous time scale $\sim 30$ s, the typical
duration of a long GRB. We will show below that powerful episodic
jets are launched in the ADAF regions, i.e. at radii larger than $r
\sim 100$, since the ejection power of magnetic blobs is relatively
suppressed in the NDAF region (see discussions below
Eqs.(\ref{powerNDAF},\ref{powerADAF})).

Now we estimate the energetics and time intervals of the magnetic
blobs. The region where a magnetic
blob occurs, i.e, the flux rope region in the disk corona, is
special because the available {\em free}
magnetic energy is large due to the topological structure of the
magnetic field. By analogy with the CME theory of solar physics, the
total available free magnetic energy of one blob is \citep{lin98}:
\be E_{\rm free}\approx 0.5\times \left(\frac{1}{12}{B_0^2}
V\right). \ee Here $B_0$ is the magnetic strength in the accretion
disk, $V\sim \pi R^3$ is the volume of the system. Defining
$\beta=P_{mag}/P_{tot}$, where
$P_{tot}=(P_{gas}+P_{rad}+P_{mag}+P_\nu) \sim P_{gas}$ in NDAF and
$P_{tot}=(P_{gas}+P_{rad}+P_{mag}) \sim P_{rad}$ in ADAF (where
$P_{mag}$($\equiv B^2/8\pi$), $P_{gas}$, $P_{rad}$, $P_\nu$ and $P_{tot}$ are magnetic,
gas, radiation, neutrino, and total pressure in the accretion disk,
respectively), one can estimate the strength of magnetic field $B_0$
for a given $\beta$. Numerical simulations show that if the $\alpha$ viscosity is intrinsically the magnetic stress associated with the MHD turbulence driven by the magnetorotational instability \citep{BH91,BH98}, as widely accepted, the values of $\alpha$ and $\beta$ are not independent \citep{blackman08}. This is confirmed by recent numerical simulations \citep{hawley2011,sorathia2012}. Noticing the different definitions between our work and \cite{blackman08}, one should roughly have $\alpha / \beta\sim 0.1$ according to \cite{blackman08}. Therefore $\alpha=0.01$ implies that the value of $\beta$ should be $\beta=0.1\beta_{-1}=0.1$. In the following we adopt $\alpha=0.01$ and $\beta=0.1$ as typical values. We note, however, that a much higher $\beta \gg 1$ can be achieved if the accretion material is already moderately magnetized at the beginning of accretion
\citep{shibata90,johansen08}.

For the inner gas-pressure-dominated NDAF and
outer radiation-pressure-dominated ADAF, one has \be
\frac{B_0^2}{8\pi}\approx 0.1\beta_{-1}P_{\rm gas}, \ee and \be
\frac{B_0^2}{8\pi}\approx 0.1 \beta_{-1}P_{\rm
rad}=0.1\beta_{-1}\frac{4\sigma}{3c}T_c^4~, \ee
respectively, so
that the available energy in the NDAF and ADAF regions are \be
E_{\rm free,NDAF}= 7.4\times 10^{48}
\alpha_{-2}^{-1.1}\beta_{-1}M_3^{1.1}\dot{M}_{32}(r/100)^{0.15}\,{\rm
erg},\ee and \be E_{\rm free,ADAF}= 8.2 \times 10^{49}
\alpha_{-2}^{-1}\beta_{-1}M_3\dot{M}_{32}(r/200)^{0.5}\,{\rm
erg},\label{Efree} \ee respectively.

The time interval between two consecutive ejections can be estimated
as the timescale to accumulate and release $E_{\rm free}$ in the
flux rope system in the corona of the accretion flow. The first
timescale is the energy accumulation time. The magnetic energy of
the flux rope system is converted from the rotational energy of the
accretion flow by Alfv\'en waves propagating along the magnetic
field lines \citep{yuan09}. The energy density of the rotational
energy is $\sim \rho v_k^2$ and the energy conversion speed is the
Alfv\'en speed $v_A\equiv B_0/\sqrt{4\pi\rho}$. The energy transfer
rate by the magnetic field is then  $\dot{E}_{\rm mag}=\rho v_k^2v_A
R^2. $ The corresponding energy transfer timescale in the NDAF and
ADAF regimes are \be t_{\rm tran,NDAF}=4.2\times
10^{-3}\alpha_{-2}^{0.1}\beta_{-1}^{0.5}M_3^{0.9}(r/100)^{1.85}
~{\rm s},\ee and \be t_{\rm tran,ADAF}=4.8\times
10^{-1}\beta_{-1}^{0.5}M_3(r/200)^{1.5} ~{\rm s},\ee respectively.

The second relevant timescale is the emergence timescale of the
magnetic field line from the disk to the corona due to Parker
instability \citep{horiuchi98}: \be t_{\rm parker,NDAF}\approx
\frac{5H}{v_A}\approx 4\alpha_{-2}^{-0.1}
\beta_{-1}^{-0.5}M_3^{1.1}(r/100)^{1.15}~{\rm s}\gg t_{\rm
tran,NDAF},\ee and \be t_{\rm parker,ADAF}\approx
\frac{5H}{v_A}\approx 3.4\beta_{-1}^{-0.5}M_3(r/200)^{1.5}~{\rm
s}\gg t_{\rm tran,ADAF},\label{tparker}\ee respectively. The time interval between two
consecutive ejections is thus $t_{\rm int}=t_{\rm tran}+t_{\rm
parker} \simeq t_{\rm parker} $ for both NDAF and ADAF. The power output from these two regions is then \be P_{\rm NDAF}=1.8\times
10^{48}\alpha_{-2}^{-1}\beta_{-1}^{1.5}\dot{M}_{32}(r/100)^{-1}
~{\rm erg\,s^{-1}},\label{powerNDAF}\ee and \be P_{\rm
ADAF}=2.4\times
10^{49}\alpha_{-2}^{-1}\beta_{-1}^{1.5}\dot{M}_{32}(r/200)^{-1}~{\rm
erg\,s^{-1}},\label{powerADAF}\ee respectively. One can see that the power output
increases significantly beyond the ``transition region'' ($r\sim
100-200$) from NDAF to ADAF. So the ADAF region is the main region
for powerful magnetic blob injection\footnote{The power output from
the innermost NDAF region with $10>r>3$ can become comparable to
that from the outer ADAF. However, the blobs emitted from these
regions are closely packed, which would likely collide below the
photosphere and would not give rise to significant variability. {In addition, this power can be consumed to penetrate the star and open a funnel for blobs ejected later from the ADAF region.}}.
From now on, we focus on the ADAF only, and neglect the
subscript ``ADAF'' in the equations.

For long GRBs, the jet-corrected total energy is of the order of
$10^{51}$ erg \citep{frail01,liang08,racusin09}. Since a GRB may be powered
by multiple collisions among blobs, each blob would carry an energy of
$\sim 10^{50}$ erg. For $r\sim 100-200$, our estimate of Eq.
(\ref{Efree}) matches this observational fact well. Comparing Eqs.
(\ref{tparker}) and (\ref{tvis}), we find that $t_{\rm int} < t_{\rm
vis}$. This suggests that the system has enough time to store
magnetic energy and to eject multiple magnetic blobs before being
accreted into the black hole.

Following \citet{uzdensky2006,uzdensky2007}, we assume that a magnetic jet
can penetrate through the star and remain intact.
For episodic jet, this is possible as long as
the time interval between two consecutive blobs is shorter than the
time for the funnel to close, which is $\sim 10~{\rm s}$
\citep{wang2007}. This condition is satisfied for our model. After
escaping the star, the magnetic blob undergoes acceleration under
its own magnetic pressure gradient \citep{tchek10}. An impulsive
magnetic blob can reach $\Gamma\sim \sigma_0^{1/3}$ quickly and then
gradually accelerate as $\Gamma \propto R^{1/3}$ until reaching
$\Gamma \sim \sigma_0$ \citep{granot11}, where $\sigma_0\equiv
\frac{B_c^2}{8\pi\rho_c c^2}$ is the initial magnetization parameter
at the base of the flow, $\rho_c$ is the density of the flux rope in
the corona, and $B_c$ is the magnetic field in the corona near the
flux rope region. Unfortunately, both $B_c$ and especially $\rho_c$
are poorly constrained by current analytical theory and numerical
simulations of accretion flows. MHD numerical simulations of both
optically thin ADAFs and standard thin disks show that the density
in the corona is strongly inhomogeneous and stratified.
A density contrast as large as $\sim$ 5 orders of magnitude
has been achieved in numerical simulations \citep{hirose2009},
which can be regarded as a lower limit due to the ``density floor''
imposed in the simulations by hand to stabilize the simulations.
Given the uncertainties, we estimate the density of corona by analogy with
the case of the Sun.  The density at the bottom of the solar corona is
about $7 \sim 12$ orders of magnitude lower than the density of the
turbulence layer of the Sun (counterpart of the accretion disk main
body). We then adopt a fiducial value for the density ratio between
the disk and the flux rope $\rho/\rho_{c}\sim 10^9$.
The magnetic field strength in the
corona ($B_c$) is not so different from that in the disk. Here we
assume that the magnetic field energy density in the flux rope is
10 times weaker than that
in the accretion flow. We then have
\be \sigma_0 \approx 3500
\left(\frac{\rho/\rho_c}{10^9}\right)\beta_{-1}\left(
\frac{r}{200}\right)^{-1}. \label{sigma0} \ee {For a higher $\beta$
value as discussed above, a smaller density constrast can be
incoporated to achieve the same $\sigma_0$ value.} The Lorentz factor of
the outflow in the emission region is $\Gamma \leq \sigma_0$
depending on the radius of the emission region, $R_{\rm GRB}$, from
the central engine. Because of the inhomogeneity of the density of
the corona, the value of $\sigma_0$ is expected to be variable for
different blobs. This implies that their collisions will be
efficient.

The initial size of each blob is of order
the size of the disk, i.e. $\Delta_0 \sim R_{\rm disk} \sim 2\times
10^{8}~{\rm cm}~ M_3 (r/200)$. The blob is initially at rest, and is
accelerated by magnetic pressure gradient. First it undergoes a
non-relativistic phase \citep{yuan09}, during which the bubble
continues to expand adiabatically. This phase lasts for a time scale
of reconnection near the base of the flux rope, i.e.
$t_{\rm rec} \sim R_{\rm disk} / v_{\rm rec} \sim 0.7~{\rm s}~
(r/200) (0.01c/v_{\rm rec})$. The expansion speed is essentially
$\sim c$ for a high-$\sigma$ bubble. So the size of the bubble grows
to $\Delta \sim c t_{\rm rec} \sim 2 \times 10^{10}~{\rm cm}$ before
entering the relativistic phase, during which the lab-frame width
essentially stops growing.

During the relativistic phase, the Lorentz factor at the distance
$R_0 \sim \Delta$ is $\Gamma(R_0)\sim \sigma_0^{1/3} \sim 15 ~
(10^9\rho/\rho_c)^{1/3}\beta_{-1}^{1/3} (r/200)^{-1/3}$. With a slow
increase $\Gamma \propto R^{1/3}$ \citep{granot11}, one would reach
the full Lorentz factor $\Gamma_{\rm full} = 3500$ at a radius
$R_{\rm full} \sim 3.8 \times 10^{17}$ cm. GRB prompt emission
likely occurs at much smaller radii where $\Gamma(R_{\rm GRB}) \ll
\sigma_0$. A plausible scenario would be the Internal
Collision-induced MAgnetic Reconnection and Turbulence (ICMART)
scenario conjectured by \citep{zhang10}. Within this scenario, most
magnetic energy is discharged in the ICMART region, so that after
dissipation $\sigma$ drops to around or below unity, and the final
bulk Lorentz factor $\Gamma \sim \Gamma(R_{\rm GRB}) \ll \sigma_0$.
For example, for $R_{\rm GRB} \sim 10^{15}$ cm, $\Gamma_{\rm GRB}
\sim 550$ for $\sigma_0 = 3500$, and $\Gamma_{\rm GRB} \sim 260$ for
$\sigma_0 = 350$.

The trigger of ICMART events is through internal collisions. It is
expected that $\sigma_0$ of the blobs may vary from case to case due
to the fluctuation of $\rho_c$ and $B_c$. The later ejected faster
blobs would inevitably catch up with the preceding slower ones. Such
collisions of magnetic blobs have been frequently observed in the
Sun \citep{gopalswamy02}, and are often invoked to explain the
bright knots or flares commonly observed in other astrophysical
systems such as  AGN jets and Crab nebula. The separation between
the blobs is $d \sim c t_{\rm int} \sim 10^{11}$ cm. The typical
collision radius is \be R_{\rm GRB} \sim \Gamma^2 c t_{\rm int}=
10^{15} \left(\frac{\Gamma}{100}\right)^2 \beta_{-1}^{-0.5}M_3
\left(\frac{r}{200}\right)^{1.5}~{\rm cm}. \label{Rem}\ee For a
typical GRB $\Gamma \geq 100$ \citep{lithwick01,liang10}, this
radius is consistent with various observational constraints that
suggest a relative large $R_{\rm GRB}$
\citep{kumar06,nishikori06,zhang09,shen09,fan10}.

The polarities of the magnetic field lines trapping the two colliding
blobs are in general opposite or have some large angles. This
greatly eases the trigger of ICMART events. The initial fast
reconnection would induce turbulence and a cascade of turbulent
reconnection \citep{zhang10,lazarian99}. A large fraction of the
magnetic energy stored in the blobs would be efficiently converted
into lepton energy and then into photon energy, giving rise to
radiatively efficient GRB emission \citep{zhang10,uzdensky11}.
The model can also account for the two variability components
as observed in GRBs \citep{gao12,vetere06}: the angular spreading time
\citep{piran99}
\be t_{\rm ang} \sim \frac{R_{\rm GRB}}{c\Gamma^2} \sim t_{\rm int}
\sim 3.4 \beta_{-1}^{-0.5}M_3\left(\frac{r}{200}\right)^{1.5}~{\rm s}~. \ee
corresponds to the time scale of the slow component, while the
fast variability is related to relativistic turbulence \citep{narayan09}.

\section{Summary and Discussion}

Observations to relatively well-studied black hole systems such as
AGNs and black hole X-ray binaries show the existence of two types
of jets, namely continuous and episodic ones. GRB observational data
require that the central engine launches a magnetically dominated
episodic jet. The traditional BZ jet model requires that the
accretion rate is highly variable or that a continuous jet is
disrupted by instabilities during propagation. In this paper we
propose an intrinsically episodic central engine model for GRBs by
invoking ejections of episodic magnetic blobs from a hyper-accretion
flow around a black hole.  Our basic calculations suggest that the
predicted energetics and timescales are consistent with the
observations. More detailed numerical simulations are called for to
validate the scenario.

This model has several appealing features. Firstly, the episodic
jets invoked in our model have obtained strong observational
supports in other black hole systems \citep*[e.g.,][and references
therein]{yuan09}. Observations show that they are more powerful than
continuous jets. Secondly, this model naturally satisfies the
observational requirement for the lack of a bright photosphere
component in the GRB spectra \citep{zhang09}, since the engine
natually launches a magnetized outflow.  Thirdly, episodic jets
intrinsically consist of individual blobs, whose collisions are
natually expected. These collisions are an important ingredient to
interpret GRB variability within the magnetically dominated model
\citep{zhang10}. Fourthly, the directions of the magnetic fields
surrounding the two adjacent blobs in general have some angles. This
greatly eases triggering fast reconnection and the subsequent
turbulence, which can efficiently convert magnetic energy into
radiation \citep{zhang10,lazarian99,mckinney11}.

We add two notes here. First, we did not discuss collimation of episodic jets in the progenitor stellar envelope. We point out that the same physics invoked in the magnetic tower model \citep{uzdensky2006} should equally apply to our model. Second, the scenario we propose should also work for a neutron star, not necessarily a black hole, as has been discussed by some previous authors \citep[e.g.][]{blackman96,kluzniak98,dai06,metzger11}.

It may be difficult to differentiate this model from other GRB central engine models from the GRB observational data. Due to their large distances and small scales, it is essentially impossible to witness ejection of magnetic blobs from GRBs. Nonetheless, observing episodic blobs may be possible for nearby AGNs and X-ray binaries with high spatial and temporal resolution observations in the future. These observations may be used to verify this generic episodic central engine model.

\acknowledgements

We thank Ramesh Narayan, Tsvi Piran, and Rongfeng Shen for valuable discussions and the referee for useful comments. FY is supported by the NSFC (grants 10825314, 10833002, 11121062, and 11133005), the National Basic Research Program of China (973 Program 2009CB824800), and the CAS/SAFEA International Partnership Program for Creative Research Teams. BZ is supported by National Science
Foundation (grant AST-0908362) and NASA (grant NNX10AD48G).

\end{document}